# A Decision Support Tool for Inferring Further Education Desires of Youth in Sri Lanka

M.F.M Firdhous*
Department of Information Technology
Faculty of Information Technology
University of Moratuwa
Moratuwa
Sri Lanka

Ravindi Jayasundara
Department of Mathematics
Faculty of Engineering
University of Moratuwa
Moratuwa
Sri Lanka

*corresponding author

*Abstract* – **This paper presents the results of a study carried out to identify the factors that influence the further education desires of Sri Lankan youth. Statistical modeling has been initially used to infer the desires of the youth and then a decision support tool has been developed based on the statistical model developed. In order to carry out the analysis and the development of the model, data collected as part of the National Youth Survey has been used. The accuracy of the model and the decision support tool has been tested by using a random data sets and the accuracy was found to be well above 80 percent, which is sufficient for any policy related decision making.**

## I. INTRODUCTION

Sri Lanka has witnessed several incidents of youth unrest in the recent past. Out of these insurgencies, two insurgencies involved the youth of the south while the other one involved the youth of the north. There have been many discussions and debates about youth unrest and the increasingly violent and intolerant nature of their conflicts. Since these discussions have been rather impressionistic there has always been the need for systematic studies to obtain information on Sri Lankan youth and their background and desires [1]. In order to collect up to date information, targeting to explore facts on Sri Lankan youth and their perceptions, an island wide national youth survey has been carried out. This has been conducted as a joint undertaking involving the United Nations Development programme (UNDP) and other six Sri Lankan and German institutions in the turn of the century. In this survey they have considered four main segments of youth, that is, their politics, conflicts, employment and education. Further Education Desires of youth have been selected to be studied further in this research. The relationship between the types of further education desire of youth in Sri Lanka had been studied with relation to other social factors.

Education domain consists of many different areas but presently in Sri Lanka only a few areas are catered by the national educational institutes [1]. By finding out the educational desires of the youth, it will be possible to design and develop educational and professional programmes and institutes which can be readily accepted by the youth and give better results than that can be achieved by only pursuing traditional programmes. Data Mining which is a powerful tool that can recognize and unearth significant facts, relationships, trends and patterns can be employed to discover this information [2]. In this project, a data mining model has been developed to predict the educational desire of youths at an early stage from other social data.

## II. THEORETICAL BACKGROUND

Descriptive statistics are used to describe the basic features of the data in a study. They provide simple summaries about the sample and the measures. Together with simple graphics analysis, they form the basis of virtually every quantitative analysis of data [3]. Univariate analysis is the simplest form of quantitative (statistical) analysis.

The analysis is carried out with the description of a single variable and its attributes of the applicable unit of analysis. Univariate analysis contrasts with bivariate analysis – the analysis of two variables simultaneously – or multivariate analysis – the analysis of multiple variables simultaneously. Univariate analysis is also used primarily for descriptive purposes, while bivariate and multivariate analysis are geared more towards explanatory purposes [4]. Univariate analysis is commonly used in the first stages of research, in analyzing the data at hand, before being supplemented by more advance, inferential bivariate or multivariate analysis [5]. Pearson's chi-square test is the best-known of several chi-square tests – statistical procedures whose results are evaluated by reference to the chi-square distribution [6].

With large samples, a chi-square test can be used. However, the significance value it provides is only an approximation, because the sampling distribution of the test statistic that is calculated is only approximately equal to the theoretical chi-squared distribution. The approximation is inadequate when sample sizes are small, or the data are very unequally distributed among the cells of the table, resulting in the cell counts predicted on the null hypothesis (the "expected values") being low. The usual rule of thumb for deciding whether the chi-squared approximation is good enough is that the chi-squared test is not suitable when the expected values in any of the cells of a contingency table are below 5 or below 10 when there is only one degree of freedom [7]. In contrast the Fisher exact test is, as its name states, exact, and it can therefore be used regardless of the sample characteristics [8]. For hand calculations, the test is only feasible in the case of a 2×2 contingency table. However the principle of the test can be extended to the general case of an m×n table [9].

Logistic regression is most frequently employed to model the relationship between a dichotomous (binary) outcome variable and a set of covariants, but with a few modifications it may also be used when the outcome variable is polytomous [10].

The extension of the model and the methods from a binary outcome variable to a polytomous outcome variable can be easily illustrated when the outcome variable has three categories. Further generalization to an outcome variable with more than three categories is more of a notation problem than a conceptual one [11]. Hence, it will be considered only the situation when the outcome variable has three categories.

Main objective of fitting this statistical model is to find out the sequence of variables being significant to the model, so that the sequence of variables, as a whole or a subsequence starting from the first variable, will be used as necessary in constructing a decision tree. In this study we make use of this statistical model not for interpretations but only for doing a comparison with the outcome of a Data Mining approach in decision making.

### III. ANALYSIS

Univariate analysis is carried out with the purpose of analyzing each variable independently from other variables. Therefore each of the categorical variables measured as a factor is cross tabularized with the dependent variable "Type of Further Educational Desires" calculating percentages of respondents belonging to each combination of levels, and the Chi-Square Test is performed in order to measure the strength of association between factors and the response of interest. A tolerance rate of 20 percent has been fixed as the significance level for further analysis. Table 1 shows the results of the analysis.

Table 1: Results of Univariate Analysis at 20% Tolerance Level

| Explanatory Variable | Pearson's Chi-square Value | Degrees of Freedom | Asymptotic Level of Significance | Significant Variables at 20% level |
|---|---|---|---|---|
| Age group | 208.0 | 6 | 0.0000 | √ |
| Gender | 68.751 | 3 | 0.0000 | √ |
| Educational level | 198.1 | 6 | 0.0000 | √ |
| Ethnicity | 53.032 | 12 | 0.0000 | √ |
| Province | 198.1 | 24 | 0.0000 | √ |
| Sector | 6.882 | 6 | 0.2020 | |
| Social class (self definition) | 77.154 | 9 | 0.0000 | √ |
| Present financial situation | 25.455 | 6 | 0.0000 | √ |
| Financial situation in past | 10.391 | 6 | 0.0810 | √ |
| Whether school provide good education | 7.723 | 9 | 0.4860 | |
| Major problems with education | 95.518 | 21 | 0.0000 | √ |
| Access to educational facilities | 90.419 | 6 | 0.0000 | √ |
| Type of activity | 1066.0 | 18 | 0.0000 | √ |

From the results shown in Table 1, the variables with their p-values less than 0.2 have been detected as significant.

#### A. Fitting a Statistical Model

The main purpose of this part of analysis is to determine the factors, which affect or are associated with having different types of Further Educational Desires in youth. Though several variables have been identified as significant factors, where each could independently build a significant effect on developing different wishes on education among youth, due to the confounding nature of these factors, it is not easy to conclude on their corporative influence on making Further Education Desires different in people.

Therefore this modeling approach can be very much useful in detecting the genuine effect of these factors when adjusted for some other factors as well.

Since the response variable is Multinomial and the scale of response levels are Nominal, it was decided to work out a "logit" link in regression modeling. Therefore a Generalized Logit Model will be fitted to accomplish the objective.

#### B. Fitting the best fitted Generalized Logit Model

The Forward Selection procedure is used in selecting variables to the model. In assessing the fit of the terms to the model, the difference in deviance of the two models compared, which is distributed as Chi-Squared has been used at the 5% significance level. However the terms will be selected to the model, as they do the best representation of all the data.

The results obtained in following the steps of fitting a Generalized Logit Model using the procedure CATMOD in SAS package, are tabularized in the body of the analysis.

Let the Null Model

be, $\log(\frac{\pi_f}{\pi_F}) = \alpha_f$;

where $\pi_f$ is the probability that a respondent has the $f^{th}$ type of Further Education Desire, $f \neq F$, and type F is the Further Education Desire category "No Desire".

Fitting Main Effects to the Model

Step 1: Null Model vs One Variable Model (Model 1)

Table 2 shows sample of data used in devising Model 1.

Table 2: Adding the 1st Most Significant Variable to the Model

| Model 1 (terms added) | Raw Deviance (-2Log likelihood) | Difference in Deviance | Difference in df | p-value |
|---|---|---|---|---|
| Null Model | 4540.468 | 0 | - | - |
| Age group | 4332.934 | 207.5343 | 4 | 9.00999E-44 |
| Gender | 4474.573 | 65.8957 | 2 | 4.90829E-15 |
| Educational level | 4359.197 | 181.2715 | 4 | 3.97611E-38 |
| Ethnicity | 4490.291 | 50.1769 | 7 | 1.33346E-08 |
| Province | 4401.342 | 139.1265 | 16 | 1.06808E-21 |
| Social class | 4477.97 | 62.4984 | 6 | 1.39659E-11 |
| Present financial situation | 4517.147 | 23.3216 | 4 | 0.000109204 |
| Financial situation in past | 4532.136 | 8.332 | 4 | 0.080146321 |
| Major problems with education | 4452.271 | 88.1975 | 14 | 8.30242E-13 |
| Access to educational facilities | 4450.442 | 90.0261 | 4 | 1.30006E-18 |
| **Type of activity** | **3450.942** | **1089.527** | **10** | **9.5521E-228** |

From Table 2, the lowest p-value is associated with the variable 'Type of activity'. The selection procedure of the most significant variable requires that 'Type of activity' be added to the Null Model as the first step of developing a model where the 'Type of Further Education Desire' being the response variable.

The explanatory variables in the model: Type of activity

Model 1: $\log(\frac{\pi_{if}}{\pi_{iF}}) = \alpha_f + x'_i \beta_f$

where $\pi_{if}$ is the probability that a respondent in 'Type of activity' category i has the $f^{th}$ type of Further Education Desire, $f \neq F$, and type F is the Further Education Desire category "No Desire".

Thus two logits are modeled for each activity type: the logit comparing Technical/Vocational Education to No Desire and the logit comparing University/Higher Education to No Desire.

*Model 1: for Type of activity i*

logit(response1/response3)i=1 models the probability of response category 1 relative to the response category 3.

logit(response2/response3)i=2 models the probability of response category 2 relative to the response category 3.

There are separate sets of intercept parameters $\alpha_f$ and regression parameter $\beta_f$ for each logit and the matrix xi is the explanatory variable for the $i^{th}$ population.

Step 2: Model 1 vs Two Variable Model (Model 2)

Table 3 shows sample of data used in devising Model 2.

Table 3: Adding the 2nd Most Significant Variable to the Model

| Model 2 (terms added) | Raw Deviance (-2Log likelihood) | Difference in Deviance | Difference in df | p-value |
|---|---|---|---|---|
| Model 1 | 3450.942 | 0 | - | - |
| Age group | 3437.718 | 13.2236 | 4 | 0.01023338 |
| Gender | 3434.514 | 16.4276 | 2 | 0.000270889 |
| **Educational level** | **3288.146** | **162.7954** | **4** | **3.67573E-34** |
| Ethnicity | 3427.465 | 23.4763 | 7 | 0.001407636 |
| Province | 3316.505 | 134.4366 | 16 | 8.79825E-21 |
| Social class | 3420.852 | 30.0896 | 6 | 3.77963E-05 |
| Present financial situation | 3432.545 | 18.3971 | 4 | 0.001031951 |
| Financial situation in past | 3448.478 | 2.4638 | 4 | 0.65112949 |
| Major problems with education | 3427.67 | 23.2713 | 14 | 0.055995893 |
| Access to educational facilities | 3437.028 | 13.9133 | 4 | 0.00757697 |

Since the variable 'Educational Level' has the lowest p-value, it was brought into the model that has been adjusted for 'Type of activity'.

The explanatory variables in the model: Type of activity, Educational Level

Model 2: $\log(\frac{\pi_{ijf}}{\pi_{ijF}}) = \alpha_f + x'_{ij} \beta_f$

where the matrix $x_{ij}$ is the set of explanatory variables for the $ij^{th}$ population.

Step 3: Model 2 vs Three Variable Model (Model 3)

Table 4 shows sample of data used in devising Model 3.

Table 4: Adding the 3rd Most Significant Variable to the Model

| Model 3 (terms added) | Raw Deviance (-2Log likelihood) | Difference in Deviance | Difference in df | p-value |
|---|---|---|---|---|
| Model 2 | 3288.146 | 0 | - | - |
| Age group | 3276.062 | 12.0842 | 4 | 0.016735994 |
| Gender | 3268.712 | 19.4339 | 2 | 6.02535E-05 |
| Ethnicity | 3267.805 | 20.3416 | 7 | 0.00487722 |
| **Province** | **3155.297** | **132.8494** | **16** | **1.79279E-20** |
| Social class | 3266.442 | 21.7041 | 6 | 0.001369785 |
| Present financial situation | 3276.61 | 11.536 | 4 | 0.02115679 |
| Financial situation in past | 3285.776 | 2.3706 | 4 | 0.667946717 |
| Major problems with education | 3267.495 | 20.6513 | 14 | 0.110907989 |
| Access to educational facilities | 3281.422 | 6.7239 | 4 | 0.151218299 |

Since the variable 'Province' has the lowest p-value, it was brought into the model already adjusted for Type of activity and Educational Level.

The explanatory variables in the model: Type of activity, Educational Level, Province

Model 3: $\log(\frac{\pi_{ijkf}}{\pi_{ijkF}}) = \alpha_f + x'_{ijk}\beta_f$

where the matrix $x_{ijk}$ is the set of explanatory variables for the $ijk^{th}$ population.

Step 4: Model 3 vs Four Variable Model (Model 4)

Table 5 shows sample of data used in devising Model 4.

Table 5: Adding the 4th Most Significant Variable to the Model

| Model 4 (terms added) | Raw Deviance (-2Log likelihood) | Difference in Deviance | Difference in df | p-value |
|---|---|---|---|---|
| Model 3 | 3155.297 | 0 | - | - |
| Age group | 3145.734 | 9.5632 | 4 | 0.048464702 |
| **Gender** | **3133.988** | **21.3087** | **2** | **2.3598E-05** |
| Ethnicity | 3152.673 | 2.6242 | 7 | 0.917458009 |
| Social class | 3128.48 | 26.8167 | 6 | 0.000156716 |
| Present financial situation | 3146.391 | 8.9055 | 4 | 0.063505426 |
| Financial situation in past | 3152.922 | 2.3747 | 4 | 0.667204138 |
| Major problems with education | 3138.381 | 16.9154 | 14 | 0.260714267 |
| Access to educational facilities | 3140.936 | 14.3609 | 4 | 0.006227982 |

In Step 4, Gender was found significant and brought into the model.

The explanatory variables in the model: Type of activity, Educational Level, Province, Gender

Model 4: $\log(\frac{\pi_{ijklf}}{\pi_{ijklF}}) = \alpha_f + x'_{ijkl}\beta_f$

where the matrix $x_{ijkl}$ is the set of explanatory variables for the $ijkl^{th}$ population.

Step 5: Model 4 vs Five Variable Model (Model 5)

Table 6 shows sample of data used in devising Model 5.

Table 6: Adding the 5th Most Significant Variable to the Model

| Model 5 (terms added) | Raw Deviance (-2Log likelihood) | Difference in Deviance | Difference in df | p-value |
|---|---|---|---|---|
| Model 4 | 3133.988 | 0 | - | - |
| Age group | 3124.457 | 9.5309 | 4 | 0.049116179 |
| Ethnicity | 3131.679 | 2.3095 | 7 | 0.940746391 |
| **Social class** | **3108.071** | **25.9174** | **6** | **0.00023067** |
| Present financial situation | 3125.265 | 8.7235 | 4 | 0.068394724 |
| Financial situation in past | 3131.218 | 2.7702 | 4 | 0.596987596 |
| Major problems with education | 3117.493 | 16.495 | 14 | 0.284088473 |
| Access to educational facilities | 3118.975 | 15.0128 | 4 | 0.004674743 |

In Step 5, 'Social class' was found significant and brought into the model.

The explanatory variables in the model: Type of activity, Educational Level, Province, Gender, Social class

Model 5: $\log(\frac{\pi_{ijklmf}}{\pi_{ijklmF}}) = \alpha_f + x'_{ijklm}\beta_f$

where the matrix $x_{ijklm}$ is the set of explanatory variables for the $ijklm^{th}$ population.

Step 6: Model 5 vs Six Variable Model (Model 6)

Table 7 shows sample of data used in devising Model 6.

Table 7: Adding the 6th Most Significant Variable to the Model

| Model 6 (terms added) | Raw Deviance (-2Log likelihood) | Difference in Deviance | Difference in df | p-value |
|---|---|---|---|---|
| Model 5 | 3133.988 | 0 | - | - |
| **Age group** | **3124.457** | **9.5309** | **4** | **0.049116179** |
| Ethnicity | 3131.679 | 2.3095 | 7 | 0.940746391 |
| Present financial situation | 3125.265 | 8.7235 | 4 | 0.068394724 |
| Financial situation in past | 3131.218 | 2.7702 | 4 | 0.596987596 |
| Major problems with education | 3117.493 | 16.495 | 14 | 0.284088473 |
| Access to educational facilities | 3118.975 | 15.0128 | 4 | 0.004674743 |

In Step 6, 'Age group' was found significant and brought into the model.

The explanatory variables in the model: Type of activity, Educational Level, Province, Gender, Social class, Age group

Model 6: $\log(\frac{\pi_{ijklmnf}}{\pi_{ijklmnF}}) = \alpha_f + x'_{ijklmn}\beta_f$

where the matrix $x_{ijklmn}$ is the set of explanatory variables for the $ijklmn^{th}$ population.

It was observed that addition of the remaining variables did not improve the results. Hence the Model 6 has been identified as the best main effect model.

*C. Improving the Model*

It was further investigated to determine if the addition of two way interaction terms improved the model. The importance of an interaction term was assessed by checking the impact of the difference in deviance of the model.

Step 7: Model 6 Vs Model 7

Table 8 shows sample of data used in devising Model 7.

Table 8: Adding the 1st Most Significant 2-Way Variable to the Model

| Model 7 (interaction term added) | Raw Deviance | Difference in Deviance | Difference in df | p-value |
|---|---|---|---|---|
| Model 6 | 3108.0707 | 0 | 0 | - |
| acti*class | 3077.8665 | 30.2042 | 23 | 0.143629043 |
| acti*age | 3073.7177 | 34.353 | 17 | 0.007557411 |
| **edu*pro** | **2975.2098** | **132.8609** | **3** | **1.30766E-28** |
| edu*gen | 3098.4356 | 9.6351 | 2 | 0.008086575 |
| pro*class | 3072.2071 | 35.8636 | 34 | 0.381101733 |
| pro*age | 3076.8779 | 31.1928 | 33 | 0.557285717 |
| gen*class | 3095.0509 | 13.0198 | 6 | 0.042722597 |
| gen*age | 3098.6612 | 9.4095 | 8 | 0.308936646 |
| class*age | 3093.4829 | 14.5878 | 16 | 0.555009994 |
| eth*mprob | 3055.5054 | 52.5653 | 45 | 0.204371948 |
| eth*faci | 3094.1994 | 13.8713 | 21 | 0.875063689 |
| fin *mprob | 3041.6852 | 66.3855 | 42 | 0.009620228 |
| fin *faci | 3079.7823 | 28.2884 | 16 | 0.029200335 |
| finp*mprob | 3056.0176 | 52.0531 | 43 | 0.162103659 |
| finp*faci | 3084.6956 | 23.3751 | 16 | 0.104069216 |
| mprob *age | 3051.2768 | 56.7939 | 42 | 0.063390185 |
| mprob*faci | 3052.4179 | 55.6528 | 44 | 0.111883885 |
| faci*gen | 3093.6093 | 14.4614 | 8 | 0.070504003 |
| faci*class | 3084.5775 | 23.4932 | 14 | 0.052702948 |
| faci*age | 3082.1476 | 25.9231 | 16 | 0.055119929 |
| faci*class | 2955.1103 | 20.0995 | 13 | 0.092757 |
| faci*age | 2944.8296 | 30.3802 | 15 | 0.010622 |

The interaction between 'Education Level' and Province has been found to have an effect on the model and hence added to the model.

The explanatory variables in the model: Type of activity, Educational Level, Province, Gender, Social class, Age group, Edu*Pro

Model 7: $\log(\frac{\pi_{ijklmnf}}{\pi_{ijklmnF}}) = \alpha_f + \mathbf{x}'_{ijklmn}\beta_f + \mathbf{y}'_{(jk)}\beta_f$

where the matrix $\mathbf{x}_{ijklmn}$ is the set of explanatory variables for the $ijklmn^{th}$ population and $\mathbf{y}_{(jk)}$ is the interaction effect of Edu*Pro.

Step 8: Model 7 Vs Model 8

Table 9 shows sample of data used in devising Model 8.

Table 9: Adding the 2nd Most Significant 2-Way Variable to the Model

| Model 8 (interaction term added) | Raw Deviance | Difference in Deviance | Difference in df | p-value |
|---|---|---|---|---|
| Model 8 | 2975.2098 | 0 | 0 | - |
| acti*age | 2942.6119 | 32.5979 | 21 | 0.05087 |
| edu*gen | 2970.8828 | 4.327 | 2 | 0.114922 |
| edu*class | 2942.4092 | 32.8006 | 3 | 3.55E-07 |
| pro*gen | 2962.5509 | 12.6589 | 11 | 0.316203 |
| pro*class | 2940.9991 | 34.2107 | 27 | 0.160007 |
| gen*class | 2960.1934 | 15.0164 | 4 | 0.004667 |
| gen*age | 2961.1957 | 14.0141 | 8 | 0.081399 |
| class*age | 2954.729 | 20.4808 | 15 | 0.154254 |
| eth*mprob | 2922.7682 | 52.4416 | 42 | 0.129663 |
| eth*faci | 2960.932 | 14.2778 | 18 | 0.710811 |
| fin *mprob | 2906.8217 | 68.3881 | 40 | 0.003421 |
| fin *faci | 2945.8832 | 29.3266 | 14 | 0.009437 |
| finp*age | 2952.1219 | 23.0879 | 15 | 0.082291 |
| **finp*mprob** | **2852.9844** | **122.2254** | **42** | **9.18E-10** |
| finp*faci | 2952.4005 | 22.8093 | 15 | 0.088274 |
| mprob *age | 2913.2762 | 61.9336 | 42 | 0.024235 |
| mprob*faci | 2923.8426 | 51.3672 | 47 | 0.306598 |

In this step, the interaction between 'Financial Situation in Past (Finp)' and 'Major Problems with Education (Mprob)' has been found to have an effect on the model and hence added to the model.

The explanatory variables in the model: Type of activity, Educational Level, Province, Gender, Social class, Age group, Edu*Pro, Finp*Mprob

Model 8:

$$\log(\frac{\pi_{ijklmnf}}{\pi_{ijklmnF}}) = \alpha_f + \mathbf{x}'_{ijklmn}\beta_f + \mathbf{y}'_{(jk)}\beta_f + \mathbf{z}'_{(op)}\beta_f$$

where the matrix $\mathbf{x}_{ijklmn}$ is the set of explanatory variables for the $ijklmn^{th}$ population, the matrix $\mathbf{y}_{(jk)}$ is the interaction effect of Edu*Pro and the matrix $\mathbf{z}_{(op)}$ is the interaction effect of Finp*Mprob.

Step 9: Model 8 Vs Model 9

Table 10 shows sample of data used in devising Model 9.

Table 10: Adding the 3rd Most Significant 2-Way Variable to the Model

| Model 9 (interaction term added) | Raw Deviance | Difference in Deviance | Difference in df | p-value |
|---|---|---|---|---|
| Model 9 | 4027.772 | | | |
| acti*age | 2909.589 | 19.7647 | 17 | 0.286441 |
| edu*gen | 2925.853 | 3.5005 | 5 | 0.623312 |
| edu*class | 2894.889 | 34.4644 | 15 | 0.002929 |
| pro*class | 2896.654 | 32.6994 | 32 | 0.432465 |
| pro*age | 2908.125 | 21.2286 | 31 | 0.905749 |
| gen*class | 2913.901 | 15.4524 | 6 | 0.017015 |
| gen*age | 2928.941 | 0.413 | 4 | 0.981399 |
| class*age | 2922.754 | 6.6001 | 12 | 0.882871 |
| eth*acti | 2899.069 | 30.2848 | 19 | 0.048284 |
| eth*edu | 2915.022 | 14.3318 | 3 | 0.002487 |
| fin*acti | 2907.075 | 22.2787 | 6 | 0.001078 |
| fin *edu | 2918.808 | 10.5456 | 11 | 0.482078 |
| fin *pro | 2892.44 | 36.9141 | 33 | 0.292752 |
| finp*acti | 2909.158 | 20.1959 | 16 | 0.211527 |
| finp*edu | 2919.167 | 10.1871 | 14 | 0.74838 |
| mprob *edu | 2892.521 | 36.8324 | 33 | 0.295967 |
| **mprob *pro** | **2788.718** | **140.6361** | **87** | **0.00024** |
| mprob *gen | 2895.19 | 34.164 | 34 | 0.459862 |
| faci*class | 2908.176 | 21.1779 | 13 | 0.069484 |
| faci*age | 2909.906 | 19.4477 | 11 | 0.053516 |

In this step, the interaction between 'Major Problems with Education (Mprob)' and 'Province (Pro)' has been found to have an effect on the model and hence added to the model.

The explanatory variables in the model: Type of activity, Educational Level, Province, Gender, Social class, Age group, Edu*Pro, Finp*Mprob, Mprob*Pro

Model 9:

$$\log(\frac{\pi_{ijklmnf}}{\pi_{ijklmnF}}) = \alpha_f + \mathbf{x}'_{ijklmn}\beta_f + \mathbf{y}'_{(jk)}\beta_f + \mathbf{z}'_{(op)}\beta_f + \mathbf{t}'_{(ko)}\beta_f$$

where the matrix $x_{ijklmn}$ is the set of explanatory variables for the ijklmnth population, the matrix $y_{(jk)}$ is the interaction effect of Edu*Pro, the matrix $z_{(op)}$ is the interaction effect of Finp*Mprob and the matrix $t_{(ko)}$ is the interaction effect of Mprob*Pro.

After Step 9 no more significant two-way terms were revealed.

### D. Checking the Adequacy of the Best 2-Way Interaction Model

Goodness of Fit: Hypothesis Testing with Deviance Statistics

H0: No lack of fit

H1: There is some lack of fit

Table 11: Goodness of Fit Testing

|  | Chi-Square | df | p-value |
|---|---|---|---|
| Deviance | 2145.881 | 2930 | 1.000 |

According to the p-value, it can be concluded that the Null Hypothesis is not rejected. Hence, the test is the proof that there is no lack of fit of the model or the model developed fits the data well.

### E. Classification Table

The classification table was used to evaluate the predictive accuracy of the regression model.

Table 12: Classification Table for Model 9

| OBSERVED | PREDICTED | | | |
|---|---|---|---|---|
|  | NO DESIRE | TECHNICAL/ VOCATIONAL EDUCATION | UNIVERSITY/ HIGHER EDUCATION | % CORRECT |
| NO DESIRE | 641 | 141 | 51 | 77.0% |
| TECHNICAL/ VOCATIONAL EDUCATION | 249 | 304 | 89 | 47.4% |
| UNIVERSITY /HIGHER EDUCATION | 46 | 45 | 519 | 85.1% |
| OVERALL PERCENTAGE | 44.9% | 23.5% | 31.6% | 70.2% |

From the results shown in Table 12, it can be seen that Model 9 has the accuracy of more than 70 percent.

### F. Implementation of Data Mining Techniques

Finally the model (Model 9) developed using Univariate analysis was used to develop a data mining model. This data mining model can predict the 'Further Educational Desire' in a youth from other attributes discussed above.

### Construction of Decision Tree

A decision tree was constructed using the attributes identified as significant in the Univariate analysis. The ordering of attributes in the decision tree was also as determined in the statistical analysis. Table 13 shows the attributes and the no. of levels of each attribute. 'Type of Further Education Desire' is the class attribute.

Table 13: List of Attributes used in the Decision Tree

| Seq. No | Attribute Name | No. of Levels |
|---|---|---|
| 1 | Type of Activity | 7 |
| 2 | Educational Level | 3 |
| 3 | Province | 9 |
| 4 | Gender | 2 |
| 5 | Social Class | 4 |
| 6 | Age group | 3 |
| 7 | Financial Situation in Past | 3 |
| 8 | Major Problems with Education | 8 |
| 9 | Type of Further Education Desire | 3 |

Figure 1 shows the portion of the decision tree thus constructed.

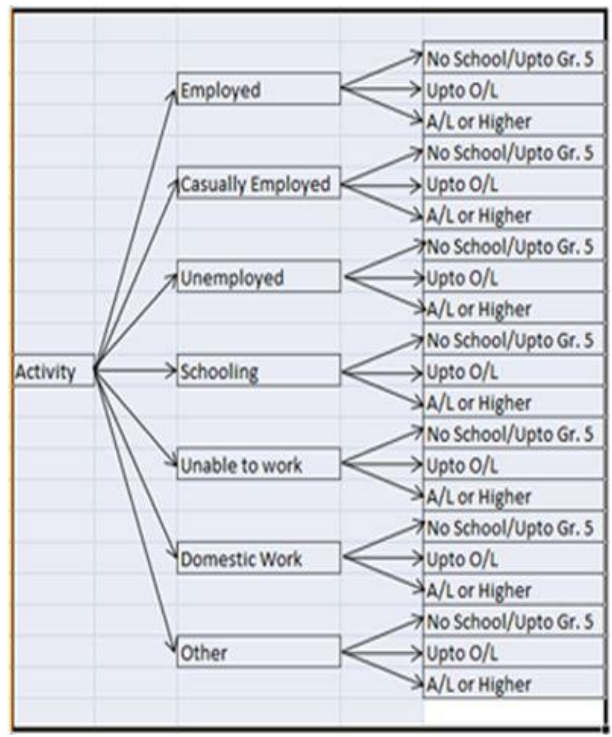

Figure 1: Portion of the Decision Tree

Figure 2 shows a sample classification rule set developed using the Decision Tree shown in Figure 1.

Rule 1: Type of Activity=Permanently Employed ^ Educational Level=No Schooling/Grade 1-5 ^ Province=Western ^ Gender=Male ^ Social Class=Middle Class ^ Age Group=20-24 yrs ⟶ No Desire

Rule 2: Type of Activity=Permanently Employed ^ Educational Level=No Schooling/Grade 1-5 ^ Province=Western ^ Gender=Male ^ Social Class=Working Class ^ Age Group=15-19 yrs ⟶ No Desire

Rule 3: Type of Activity=Permanently Employed ^ Educational Level=No Schooling/Grade 1-5 ^ Province=Central ^ Gender=Male ^ Social Class=Working Class ^ Age Group=20-24 yrs ⟶ No Desire

Rule 4: Type of Activity=Permanently Employed ^ Educational Level=No Schooling/Grade 1-5 ^ Province=Central ^ Gender=Female ^ Social Class=Working Class ^ Age Group=20-24 yrs ⟶ No Desire

Rule 5: Type of Activity=Permanently Employed ^ Educational Level=No Schooling/Grade 1-5 ^ Province=Uva ^ Gender=Male ^ Social Class=Working Class ^ Age Group=24-29 yrs ⟶ Technical/Vocational Education

Figure 2: Sample Rule Set

*G. Implementation*

A software tool was developed using Visual Basic to implement the rule set developed above. Figure 3 shows the interface of the software tool developed.

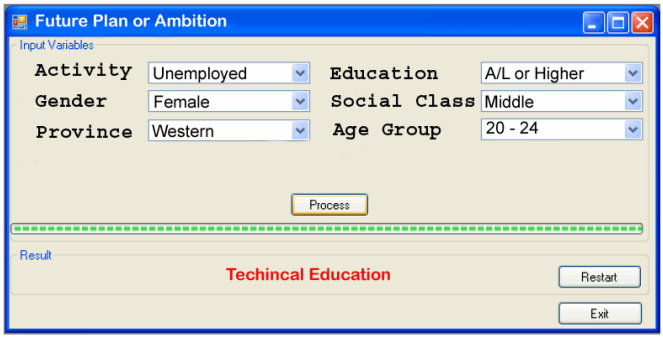

Figure 3: Interface of the Software Tool Developed

*H. Evaluation*

The system developed was tested using a random set of data containing 485 test records. The selected random set was divided into four test sets each test set containing around 125 records. Table 14 shows the classification table obtained by inputting data into the application.

Table 14: Classification Table Obtained from Test

| **OBSERVED** | **PREDICTED** ||||||||||||
|---|---|---|---|---|---|---|---|---|---|---|---|---|
| | TECHNICAL / VOCATIONAL EDUCATION ||| UNIVERSITY/ HIGHER EDUCATION |||| NO DESIRE ||||
| TECHNICAL/ VOCATIONAL EDUCATION | 21 | 13 | 19 | 22 | 3 | 5 | 5 | 7 | 13 | 17 | 16 | 7 |
| UNIVERSITY /HIGHER EDUCATION | 0 | 4 | 3 | 7 | 30 | 23 | 35 | 26 | 3 | 5 | 2 | 2 |
| NO DESIRE | 10 | 5 | 11 | 8 | 4 | 7 | 2 | 0 | 32 | 51 | 27 | 40 |

Table 15 shows the Confusion Matrices for various adjusted outcomes.

Table 15 (a): Confusion Matrix Adjusted for No Desire

| | | | Predicted ||
|---|---|---|---|---|
| | | | No Desire | Have a Desire |
| Observed | Data Set 1 | No Desire | 36 | 14 |
| | | Have a Desire | 16 | 54 |
| | Data Set 2 | No Desire | 51 | 12 |
| | | Have a Desire | 22 | 45 |
| | Data Set 3 | No Desire | 27 | 13 |
| | | Have a Desire | 18 | 62 |
| | Data Set 4 | No Desire | 40 | 8 |
| | | Have a Desire | 9 | 62 |

Table 15 (b): Confusion Matrix Adjusted for Uni/Higher Education

| | | | Predicted ||
|---|---|---|---|---|
| | | | Uni/Higher | Not So |
| Observed | Data Set 1 | Uni/Higher | 30 | 3 |
| | | Not So | 7 | 76 |
| | Data Set 2 | Uni/Higher | 23 | 9 |
| | | Not So | 12 | 86 |
| | Data Set 3 | Uni/Higher | 35 | 5 |
| | | Not So | 7 | 73 |
| | Data Set 4 | Uni/Higher | 26 | 9 |
| | | Not So | 7 | 77 |

Table 15 (c): Confusion Matrix Adjusted for Tech/Voc. Education

| | | | Predicted ||
|---|---|---|---|---|
| | | | Tech/Voc | Not So |
| Observed | Data Set 1 | Tech/Voc | 21 | 16 |
| | | Not So | 10 | 69 |
| | Data Set 2 | Tech/Voc | 13 | 22 |
| | | Not So | 9 | 86 |
| | Data Set 3 | Tech/Voc | 19 | 21 |
| | | Not So | 14 | 66 |
| | Data Set 4 | Tech/Voc | 22 | 14 |
| | | Not So | 15 | 68 |

Table 16 shows the resulting measures obtained from the tests.

Table 16: Measures Obtained from Tests

| | | Tech/Voc | Uni/High | No Desire | Avg. |
|---|---|---|---|---|---|
| Data Set 1 | TPR | 0.567568 | 0.909091 | 0.72 | 0.7322 |
| | FPR | 0.126582 | 0.084337 | 0.228571 | 0.1464 |
| | Accuracy | 0.775862 | 0.913793 | 0.75 | 0.8132 |
| Data Set 2 | TPR | 0.371429 | 0.71875 | 0.809524 | 0.6332 |
| | FPR | 0.094737 | 0.122449 | 0.328358 | 0.1818 |
| | Accuracy | 0.761538 | 0.838462 | 0.738462 | 0.7794 |
| Data Set 3 | TPR | 0.475 | 0.875 | 0.675 | 0.675 |
| | FPR | 0.175 | 0.0875 | 0.225 | 0.1625 |
| | Accuracy | 0.708333 | 0.9 | 0.741667 | 0.7833 |
| Data Set 4 | TPR | 0.611111 | 0.742857 | 0.833333 | 0.7291 |
| | FPR | 0.180723 | 0.083333 | 0.126761 | 0.1302 |
| | Accuracy | 0.756303 | 0.865546 | 0.857143 | 0.8263 |
| Overall | TPR | 0.50628 | 0.81142 | 0.75946 | 0.6923 |
| | FPR | 0.14426 | 0.0944 | 0.22717 | 0.1552 |
| | Accuracy | 0.75051 | 0.87945 | 0.77182 | 0.8005 |

From Table 16, it can be seen that the overall accuracy of the system is above 80 percent. Figure 7 shows the Receiver-Operator Characteristic (ROC) curve drawn from the data in Figure 4.

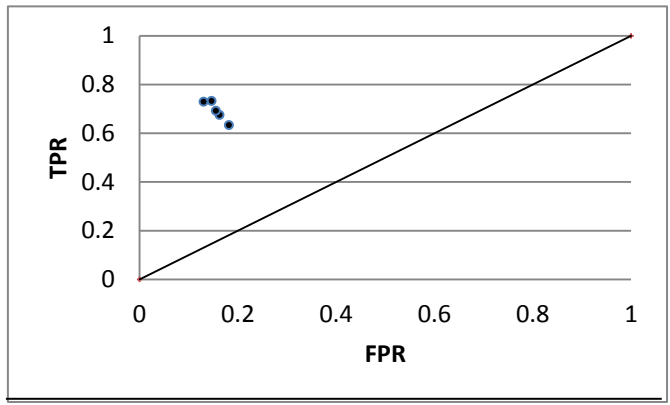

Figure 4: Receiver Operator Characteristic Curve for Model

From the ROC curve, it can be seen that the results concentrate towards the upper left hand corner. This is the proof that the accuracy of the Data Mining model is acceptable as pure random guess would lie along the diagonal line.

## IV. CONCLUSIONS

This paper presented the results of the research carried out to find the factors on which the Educational Desires of Sri Lankan Youth depends. The research found that the Educational Desires of the Youth could be predicated through the combination of several social factors. The findings of the research were finally used to design a data mining model for the predication of the Educational Desires of the Youth. This model can be used by decision makers in dealing with issues concerning youth especially their further educational requirements.

## AUTHORS PROFILE

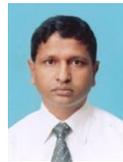

Mohamed Fazil Mohamed Firdhous is a senior lecturer attached to the Faculty of Information Technology of the University of Moratuwa, Sri Lanka. He received his BSc Eng., MSc and MBA degrees from the University of Moratuwa, Sri Lanka, Nanyang Technological University, Singapore and University of Colombo Sri Lanka respectively. In addition to his academic qualifications, he is a Chartered Engineer and a Corporate Member of the Institution of Engineers, Sri Lanka, the Institution of Engineering and Technology, United Kingdom and the International Association of Engineers. Mohamed Firdhous has several years of industry, academic and research experience in Sri Lanka, Singapore and the United States of America.

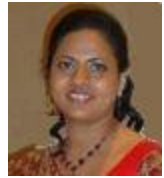

Ravindi Jayasundara is a lecturer attached to the Department of Mathematics, Faculty of Engineering of the University of Sri Lanka. She completed the BSc (Hons) degree in Statistics with Second Class (Upper Division) at the University of Colombo, Sri Lanka and MSc in Operations Research at the University of Moratuwa, Sri Lanka. Ravindi's research interests are in statistics, operations research and data mining.